\documentclass[traditabstract]{aa} 
\usepackage{graphicx}
\usepackage{textcomp}
\usepackage{amsmath}
\usepackage{upgreek}
\usepackage{subfigure}
\usepackage{txfonts}
\usepackage{natbib}
\begin{document}

   \title{Calibrating High-Precision Faraday Rotation Measurements for
     LOFAR and the Next Generation of Low-Frequency Radio Telescopes}

   \author{C. Sotomayor-Beltran\inst{1}
		\and
		C.~Sobey\inst{2}\and
		J.~W.~T.~Hessels\inst{3,4}\and
		G.~de~Bruyn\inst{3,5}\and
		A.~Noutsos\inst{2}\and
		A.~Alexov\inst{4,6}\and
		J.~Anderson\inst{2}\and
		A.~Asgekar\inst{3}\and
		I.~M.~Avruch\inst{7,5,3}\and
		R.~Beck\inst{2}\and
		M.~E.~Bell\inst{8,9}\and
		M.~R.~Bell\inst{10}\and
		M.~J.~Bentum\inst{3}\and
		G.~Bernardi\inst{5}\and
		P.~Best\inst{11}\and
		L.~Birzan\inst{12}\and
		A.~Bonafede\inst{13,14}\and
		F.~Breitling\inst{15}\and
		J.~Broderick\inst{9}\and
		W.~N.~Brouw\inst{3,5}\and
		M.~Br\"uggen\inst{13}\and
		B.~Ciardi\inst{10}\and
		F.~de Gasperin\inst{13,10}\and
		R.-J.~Dettmar\inst{1}\and
		A.~van Duin\inst{3}\and
		S.~Duscha\inst{3}\and
		J.~Eisl\"offel\inst{16}\and
		H.~Falcke\inst{17,3}\and
		R.~A.~Fallows\inst{3}\and
		R.~Fender\inst{9}\and
		C.~Ferrari\inst{18}\and
		W.~Frieswijk\inst{3}\and
		M.~A.~Garrett\inst{3,12}\and
		J.~Grie\ss{}meier\inst{3,19}\and
		T.~Grit\inst{3}\and
		A.~W.~Gunst\inst{3}\and
		T.~E.~Hassall\inst{9,20}\and
		G.~Heald\inst{3}\and
		M.~Hoeft\inst{16}\and
		A.~Horneffer\inst{2}\and
		M.~Iacobelli\inst{12}\and
		E.~Juette\inst{1}\and
		A.~Karastergiou\inst{21}\and
		E.~Keane\inst{2}\and
		J.~Kohler\inst{2}\and
		M.~Kramer\inst{2,20}\and
		V.~I.~Kondratiev\inst{3,22}\and
		L.~V.~E.~Koopmans\inst{5}\and
		M.~Kuniyoshi\inst{2}\and
		G.~Kuper\inst{3}\and
		J.~van Leeuwen\inst{3,4}\and
		P.~Maat\inst{3}\and
		G.~Macario\inst{18}\and
		S.~Markoff\inst{4}\and
		J.~P.~McKean\inst{3}\and
		D.~D.~Mulcahy\inst{2}\and
		H.~Munk\inst{3}\and
		E.~Orru\inst{3,17}\and
		H.~Paas\inst{23}\and
		M.~Pandey-Pommier\inst{12,24}\and
		M.~Pilia\inst{3}\and
		R.~Pizzo\inst{3}\and
		A.~G.~Polatidis\inst{3}\and
		W.~Reich\inst{2}\and
		H.~R\"ottgering\inst{12}\and
		M.~Serylak\inst{19,25}\and
		J.~Sluman\inst{3}\and
		B.~W.~Stappers\inst{20}\and
		M.~Tagger\inst{19}\and
		Y.~Tang\inst{3}\and
		C.~Tasse\inst{26}\and
		S.~ter~Veen\inst{17}\and
		R.~Vermeulen\inst{3}\and
		R.~J.~van Weeren\inst{27,12,3}\and
		R.~A.~M.~J.~Wijers\inst{4}\and
		S.~J.~Wijnholds\inst{3}\and
		M.~W.~Wise\inst{3,4}\and
		O.~Wucknitz\inst{28,2}\and
		S.~Yatawatta\inst{3}\and
		P.~Zarka\inst{26}
          }

   \institute{Astronomisches Institut der Ruhr-Universit\"at Bochum, Universit\"atsstr. 150, 44780 Bochum, Germany\\
              \email{sotomayor@astro.rub.de}
		\and
		{Max-Planck-Institut f\"{u}r Radioastronomie, Auf dem H\"ugel 69, 53121 Bonn, Germany}
 		\and
		{ASTRON, the Netherlands Institute for Radio Astronomy, Postbus 2, 7990 AA, Dwingeloo, The Netherlands}
		\and 
		{Astronomical Institute ``Anton Pannekoek,'' University of Amsterdam, Science Park 904, 1098 XH Amsterdam, The Netherlands}
		\and 
		{Kapteyn Astronomical Institute, PO Box 800, 9700 AV Groningen, The Netherlands}
		\and 
		{Space Telescope Science Institute, 3700 San Martin Drive, Baltimore, MD 21218, USA}
		\and
		{SRON Netherlands Insitute for Space Research, Sorbonnelaan 2, 3584 CA, Utrecht, The Netherlands}
		\and 
		{ARC Centre of Excellence for All-sky astrophysics (CAASTRO), Sydney Institute of Astronomy, University of Sydney Australia}
		\and 
		{School of Physics and Astronomy, University of Southampton, Southampton, SO17 1BJ, UK}
		\and 
		{Max Planck Institute for Astrophysics, Karl Schwarzschild Str. 1, 85741 Garching, Germany}
		\and 
		{Institute for Astronomy, University of Edinburgh, Royal Observatory of Edinburgh, Blackford Hill, Edinburgh EH9 3HJ, UK}
		\and 
		{Leiden Observatory, Leiden University, PO Box 9513, 2300 RA Leiden, The Netherlands}
		\and 
		{University of Hamburg, Gojenbergsweg 112, 21029 Hamburg, Germany}
		\and 
		{Jacobs University Bremen, Campus Ring 1, 28759 Bremen, Germany}
		\and 
		{Leibniz-Institut f\"{u}r Astrophysik Potsdam (AIP), An der Sternwarte 16, 14482 Potsdam, Germany}
		\and 
		{Th\"{u}ringer Landessternwarte, Sternwarte 5, D-07778 Tautenburg, Germany}
		\and 
		{Department of Astrophysics/IMAPP, Radboud University Nijmegen, P.O. Box 9010, 6500 GL Nijmegen, The Netherlands}
		\and 
		{Laboratoire Lagrange, UMR7293, Universit\'{e} de Nice Sophia-Antipolis, CNRS, Observatoire de la C\'{o}te d'Azur, 06300 Nice, France}
		\and 
		{Laboratoire de Physique et Chimie de l'Environnement et de l'Espace, LPC2E UMR 7328 CNRS, 45071 Orl\'{e}ans Cedex 02, France}
		\and 
		{Jodrell Bank Center for Astrophysics, School of Physics and Astronomy, The University of Manchester, Manchester M13 9PL,UK}
		\and 
		{Astrophysics, University of Oxford, Denys Wilkinson Building, Keble Road, Oxford OX1 3RH}
		\and 
		{Astro Space Center of the Lebedev Physical Institute, Profsoyuznaya str. 84/32, Moscow 117997, Russia}
		\and 
		{Center for Information Technology (CIT), University of Groningen, The Netherlands}
		\and
		{Centre de Recherche Astrophysique de Lyon, Observatoire de Lyon, 9 av Charles Andr\'{e}, 69561 Saint Genis Laval Cedex, France}
		\and 
		{Station de Radioastronomie de Nan\c{c}ay, Observatoire de Paris, CNRS/INSU, 18330 Nan\c{c}ay, France}
		\and
		{LESIA, UMR CNRS 8109, Observatoire de Paris, 92195   Meudon, France}
		\and 
		{Harvard-Smithsonian Center for Astrophysics, 60 Garden Street, Cambridge, MA 02138, USA}
		\and 
		{Argelander-Institut f\"{u}r Astronomie, University of Bonn, Auf dem H\"{u}gel 71, 53121, Bonn, Germany} 
             }

  \date{Received 13 November 2012 / Accepted 25 February 2013}
 
  \abstract{

    Faraday rotation measurements using the current and next
    generation of low-frequency radio telescopes will provide a
    powerful probe of astronomical magnetic fields.  However,
    achieving the full potential of these measurements requires
    accurate removal of the time-variable ionospheric Faraday rotation
    contribution.  We present {\tt ionFR}, a code that calculates the
    amount of ionospheric Faraday rotation for a specific epoch,
    geographic location, and line-of-sight.
    {\tt ionFR} uses a number of publicly available, GPS-derived total
    electron content maps and the most recent release of the
    International Geomagnetic Reference Field.  We describe
    applications of this code for the calibration of radio
    polarimetric observations, and demonstrate the high accuracy of
    its modeled ionospheric Faraday rotations using LOFAR pulsar
    observations.  These show that we can accurately determine
    some of the highest-precision pulsar rotation measures ever
    achieved.  Precision rotation measures can be used to monitor
    rotation measure variations --- either intrinsic or due to the
    changing line-of-sight through the interstellar medium. This
    calibration is particularly important for nearby sources,
    where the ionosphere can contribute a significant fraction of the
    observed rotation measure.  We also discuss planned improvements
    to {\tt ionFR}, as well as the importance of ionospheric Faraday
    rotation calibration for the emerging generation of low-frequency
    radio telescopes, such as the SKA and its pathfinders.  
	}
   

   \titlerunning{Ionospheric Faraday rotation}

   \maketitle

\section{Introduction}

In recent years, low-frequency ($\nu_{\rm obs} < 300$ MHz) radio
astronomy has undergone a revival due to the construction of the
  Giant Metrewave Radio Telescope
  \citep[GMRT;]{1991ASPC...19..376S} and modern aperture array
radio telescopes such as the LOw Frequency ARray (LOFAR; Stappers et
al. 2011; van Haarlem et al. 2013, submitted), the Long-Wavelength
Array \citep[LWA;]{2010amos.confE..59K}, and the Murchison
Widefield Array \citep[MWA;]{2010rfim.workE..16M}.  The first phase
of the Square Kilometre Array \citep[SKA;]{2010iska.meetE..18G} is
also planned to feature a large number of low-frequency antennas,
operating at $\sim70-450$ MHz.  These telescopes open new scientific
horizons in the area of low-frequency radio astronomy, including the
determination of high-precision Faraday rotation measures (RMs).

Faraday rotation causes the intrinsic polarization angle
($\chi_{\mathrm{0}}$) of a signal to rotate as it propagates through a
magneto-ionic medium\footnote{A magneto-ionic medium is made up of
  ionized gas and magnetic fields (e.g., the ionosphere).}.  The
observed polarization angle of a point source can be defined as\footnote{First-order 
approximation for high frequencies. In the 
low-frequency band additional terms, not shown, may also become significant.}:
\begin{equation}
\label{eq:IONFR}
\chi = \chi_{\mathrm{0}} + (\phi_{\mathrm{ion}}+\phi_{\mathrm{ISM}}+\phi_{\mathrm{IGM}})\lambda^{\mathrm{2}},
\end{equation}
where $\chi$ denotes the observed polarization angle in radians and
$\lambda$ the observing wavelength in meters. $\phi$ is the Faraday
depth in rad m$^{\mathrm{-2}}$ of a given intervening magneto-ionic
medium.  In Eq.~\ref{eq:IONFR} three intervening media produce Faraday
rotation: the ionosphere (ion), the Galactic interstellar medium (ISM), and the
inter-galactic medium (IGM; assuming the source is extra-Galactic). 
Low-frequency observations are particularly affected because Faraday
rotation increases quadratically with wavelength.

For the case of a single polarized source positioned behind one or
more magneto-ionic media that are not emitting polarised
radiation, the Faraday depth of the source is equivalent to its RM
\citep[e.g.,]{1998MNRAS.299..189S}.  Nonetheless, here we will use
the more generic term Faraday depth. Following
\citet{2005A&A...441.1217B} we define Faraday depth as:
\begin{equation}
\label{eq:RM}
\phi(\vec{\mathrm{l}}) = 0.81 \int^{\mathrm{observer}}_{\mathrm{source}} n_{\mathrm{e}} \vec{\mathrm{B}}\cdot \mathrm{d}\vec{\mathrm{l}}\,\mathrm{rad}\,\mathrm{m}^{\mathrm{-2}},
\end{equation}
where $n_{\mathrm{e}}$ and $\vec{\mathrm{B}}$ are the electron
density (cm$^{-3}$) and magnetic field ($\upmu$G) integrated 
along the line-of-sight (LOS) to the source and 
d$\vec{\mathrm{l}}$ is the infinitesimal path length in pc.
A magnetic field pointing towards/away from the observer gives 
a positive/negative Faraday depth.

The electron density in the ionosphere dictates the lowest frequency
observable from the ground, approximately 10 MHz.  Above this
frequency, the ionosphere affects signals in three main ways:
i) differential phase delays, ii) Faraday rotation, and iii) absorption in
  the High Frequency band (HF; 3$-$30 MHz) and the low-end of the
  Very High Frequency band (VHF; 30$-$300 MHz) due to the
  presence of the so-called ``D-region'' in the daytime.
Assuming a typical observing
frequency of 150 MHz (LOFAR high band) and an ionospheric Faraday
depth of 1 rad m$^{-2}$, the additional rotation of the polarization
angle imparted by the ionosphere will be $\sim 228.9^{\circ}$.  
Although the rotation of the polarization angle
is less pronounced at higher frequencies, the Faraday depth of the
source will still be systematically affected by the ionosphere.  Due
to the direction of the geomagnetic field, ionospheric Faraday
rotation has a positive or negative contribution to the total Faraday
depth of a source when observing from the northern or southern
hemispheres, respectively.  For instance, the contribution from the
ionosphere should be corrected for in order to derive reliable Faraday
depths due to the ISM alone when observing Galactic pulsars.  This is
particularly important for pulsars that are relatively nearby and/or
located above the Galactic plane because the magnitude of the
ionospheric Faraday depth can be a significant fraction of, or even
greater than, the total observed Faraday depth.

Calibrating for ionospheric Faraday rotation is complicated because
the free electron content of the ionosphere varies depending on the
time of day, season, level of solar activity, and LOS.  The ionosphere
changes on timescales that are often shorter than the length of an
observation; e.g., \citet{2008A&A...489...69B} and
\citet{2011A&A...525A.104P} report Faraday depth variations in
polarized point sources of a few rad m$^{\mathrm{-2}}$ in 12 hours.
Calibrating for the ionosphere is therefore critical for comparing
Faraday depths of the same source at multiple epochs.  The
time-dependence of ionospheric Faraday rotation can wash out the
linear polarization when averaging over multi-hour time intervals at
long wavelengths.

Faraday depth measurements can be used to map the structure of the
Galactic magnetic field (GMF) using pulsars
\citep[e.g.,]{2006ApJ...642..868H,2008MNRAS.386.1881N} and
extragalactic sources
\citep{2007ApJ...663..258B,2011ApJ...728...97V}.  Knowledge of the
magnitude and structure of the GMF is key for understanding deflection
of high-energy cosmic rays, star-forming regions,
instability-generated turbulence, pressure on ionized gas and the
transport of heat, angular momentum and energy from cosmic rays.
Monitoring Faraday rotation variations over time also yields insights
into the polarization modes of pulsar emission and ISM magnetic field
variations \citep[e.g.,]{2004ApJS..150..317W}.  The magnetic fields
of other galaxies have also been the subject of thorough
investigation, although weak magnetic fields best detected at low
frequencies remain relatively unexplored
\citep{2009RMxAC..36....1B}.  Detecting weak, coherent magnetic
fields will provide insight into how distant cosmic-rays originating
in the discs of galaxies propagate within the halos and possibly into
intergalactic space.  Furthermore, a deeper understanding of magnetic
fields in galaxy cluster halos and relics can also be gained.
Planetary observations can be used to determine ionospheric
Faraday depths, e.g., by observing and studying the nature of Jupiter
bursts at $\sim20$ MHz \citep{2007A&A...471.1099N}.  Lastly,
  Faraday depth measurements may even provide a test for the nature of
  the accretion flow of the supermassive black hole at the Galactic
  Centre \citep[e.g.,]{2011MNRAS.415.1228P}.

Here we present {\tt ionFR}, a code that models the ionospheric
Faraday depth using publicly available global TEC maps and the latest
geomagnetic field model.  We demonstrate the robustness of {\tt ionFR}
by comparing its modeled Faraday depths with the measured Faraday
depths of pulsars from four LOFAR observing campaigns, as well as an
observation of the pulsar PSR B1937+21 with the Westerbork Synthesis
Radio Telescope (WSRT).  \S2 describes previous studies and the
methodology of the {\tt ionFR} code.  \S3 presents modeled ionospheric
Faraday depths that the software produces for varying levels of solar
activity, including for the proposed SKA sites.  \S4 briefly
introduces RM-synthesis, which was used to determine Faraday depths
here.  Comparisons between the {\tt ionFR}-modeled ionospheric Faraday
depths and the observations are shown in \S5. \S6 discusses these
results in more detail.  Conclusions are presented in \S7.

\section{Modeling the ionosphere with the {\tt ionFR} code}

Here we describe the theoretical background and methodology of {\tt
  ionFR}, which is written almost entirely in Python and is freely
available to the community via {\tt sourceforge}\footnote{{\tt
    http://sourceforge.net/projects/ionfarrot/}}.  The code returns a
table containing the ionospheric TEC\footnote{Typically measured in
TEC units (TECU)}, magnetic field
magnitude and RM along the requested LOS (see Fig.~\ref{ionPredict}
for two examples).  The required input arguments are: right
ascension and declination of the source, geographic coordinates of the
observing site, the date of observation and the type of TEC map to be
used.

\subsection{Previous implementations}

There have been several ionospheric Faraday rotation models previously
presented.  \citet{2001A&A...366.1071E} constructed a simple model
based on Global Positioning System (GPS) data.  However, this model
requires local GPS data, i.e., dual frequency GPS receivers installed at the
telescope site(s).  \citet{2008SPIE.6936E..65A} used a
well-established empirical ionospheric model, the International
Reference Ionosphere (IRI).  The IRI provides ionospheric parameters
derived mostly from ground-based instruments (e.g., ionosondes) and
some space-based instruments \citep{2008AdSpR..42..599B}.  The
model was compared solely with GPS data 
and no comparison with radio astronomical data was
presented.  The {\tt ionFR} software presented here is somewhat
similar to the {\tt TECOR} task from the Astronomical Imaging
Processing System \citep[AIPS;]{2003ASSL..285..109G}.  However,
while {\tt TECOR} requires exporting interferometric data into AIPS,
{\tt ionFR} is a standalone package.

\subsection{Ionospheric piercing point}\label{Ionopp}

\begin{figure}
\resizebox{\hsize}{!}{\includegraphics{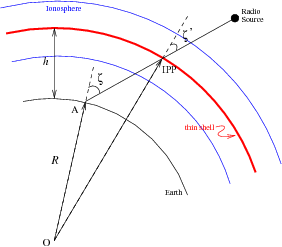}}  
\caption{Schematic representation of an astronomical signal piercing the ionosphere. The solid blue curves delineate the ionosphere. The thick red curve is the ionospheric thin shell approximation, and the thin black curve delineates the surface of the Earth.}
\label{ippoint}
\end{figure}

To calculate the ionospheric Faraday depth along the LOS, we assume a
thin spherical shell surrounding the Earth (Fig.~\ref{ippoint}). The
ionospheric Faraday depth is then calculated at the \emph{ionospheric
  piercing point} (IPP). Under these assumptions and from
Eq.~\ref{eq:RM}, the ionospheric Faraday depth is defined as:
\begin{equation}
\label{ionFD}
\phi_{\mathrm{ion}} = 2.6\times10^{-17}\,\mathrm{TEC}_{\rm LOS}\,
\mathrm{B}_{\rm LOS}\,\mathrm{rad}\, \mathrm{m}^{-2},
\end{equation}
where TEC$_{\rm LOS}$ is the total electron content at the geographic coordinates of the IPP.
TEC$_{\rm LOS}$ is defined as:
\begin{equation}
\label{ionFD}
\mathrm{TEC}_{\rm LOS}=  \int n_{\mathrm{e}} \mathrm{d}\vec{\mathrm{l}}\,\mathrm{m}^{-2}.
\end{equation}
B$_{\rm LOS}$ is the geomagnetic field intensity in gauss at the IPP. 
To facilitate the estimation of the IPPs, the code assumes that the Earth is a sphere 
of radius $R= 6371$ km.

Given the triangle defined by the points A, O, and IPP (Fig.~\ref{ippoint}), the value of the zenith angle ($\zeta^{\prime}$)
at the IPP is derived using the law of sines:
\begin{equation}
\label{zenithipp}
\mathrm{sin}(\zeta^{\prime}) = \frac{R}{R+h}\mathrm{sin}(\zeta),
\end{equation}
where $h$ is the altitude of the ionospheric thin shell, as specified in the 
ionospheric electron density files described in \S\ref{Ionoed}.
In a similar
fashion, the other geographic and topographic parameters at the IPP can
be calculated.  Spherical trigonometry is used to
calculate the latitude, longitude, and azimuth ($\varphi^{\prime}$) of
the IPP.  

\subsection{Ionospheric electron density}\label{Ionoed}

\begin{figure*}
\centering 
\includegraphics[trim = 35mm 15mm 35mm 15mm,clip=True,width=18cm]{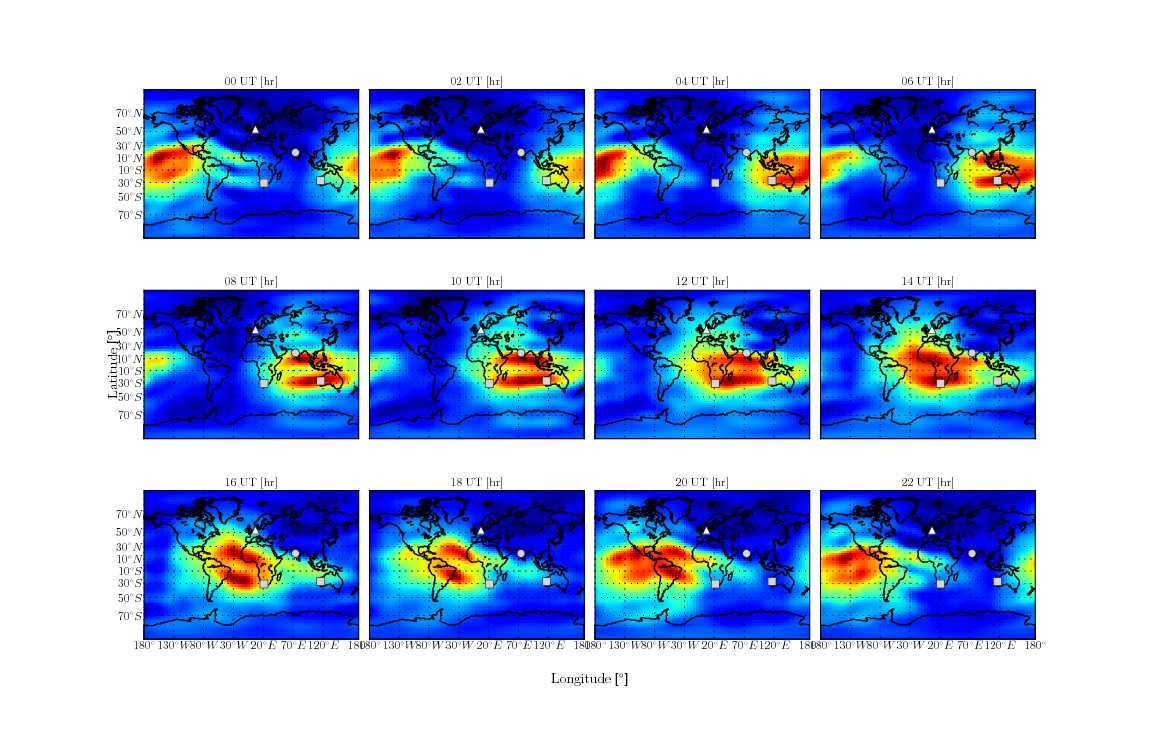} 
\caption{GIMs representing the VTEC across the globe for April 11th,
  2011 (the date of the first LOFAR observing campaign, see \S5)
  obtained courtesy of CODE. The maps range from minimum (blue) to
  maximum (red) VTEC values of $0.0 - 87.2$ TECU (1 TECU =
    10$^{16}$ electrons/m$^{2}$).  The triangles indicate the
  location of the LOFAR core stations in the Netherlands, the squares mark the SKA core sites in South Africa and Western
  Australia, and the circles indicate the site of the GMRT.}
\label{TECmap}
\end{figure*}

Measurements of the ionospheric free electron content come from
several sources.  For example, the Center for Orbit Determination in
Europe (CODE) offers global ionosphere maps (GIMs) in IONosphere map
EXchange format (IONEX), available via anonymous {\tt
  ftp}\footnote{{\tt ftp://ftp.unibe.ch/aiub/CODE/}}. IONEX files from
 CODE are derived from $\sim$ 200 GPS sites of the International Global Navigation
 Satellite System Service (IGS) and other institutions.
Figure~\ref{TECmap} illustrates twelve maps obtained from the CODE
IONEX file for April 11th, 2011.

The IONEX files provide vertical total electron content (VTEC) values
in a geographic grid ($\Delta_{\mathrm{lon.}}$ = 5$^\circ$,
$\Delta_{\mathrm{lat.}}$ = 2.5$^\circ$). VTEC is defined as the
integral of free electrons in the ionosphere along the zenith ($\zeta
= 0^{\circ}$) direction.  The time resolution of the IONEX files
provided by CODE is 2 hours.  To increase the time resolution, {\tt
  ionFR} creates new GIMs for every hour by using an interpolation
scheme that takes the rotation of the Earth into consideration
\citep[Eq.~3]{1998IGS.SCH..233}.  The software interpolates the
positional measurement grid to estimate the VTEC values at a given IPP
for each hourly GIM. The interpolation uses a 4-point formula as in
Fig.~1 of \citet[]{1998IGS.SCH..233}.

The final step in calculating the line-of-sight TEC is converting the
VTEC to the slant TEC (TEC$_{\rm LOS}$) as follows:
\begin{equation}
\label{slantTEC}
\mathrm{TEC}_{\rm LOS} = \frac{\mathrm{VTEC}}{\rm cos(\zeta^{\prime})}.
\end{equation}
CODE also provides root-mean-square (RMS) VTEC maps that are
geographically gridded in the same way as the VTEC maps.  
The uncertainties are calculated from these maps using
Eq.~\ref{slantTEC}. The 1-$\sigma$ uncertainties in the RMS VTEC maps
are typically between 2 -- 5 TECU.

\begin{figure}
\resizebox{\hsize}{!}{\includegraphics{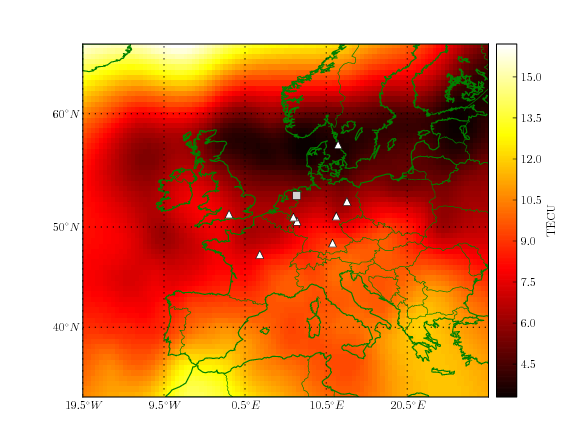}}  

\caption{The VTEC across Europe for March 23rd, 2012 (the date of the
  third LOFAR campaign, see \S5) at 00:00 UT, obtained courtesy of
  ROB. The square indicates the LOFAR core stations and the
  triangles represent the locations of the international stations.}

\label{euTECmap}
\end{figure}

{\tt ionFR} will be regularly updated to allow a greater selection of
TEC map sources.  Maps with higher spatial and temporal resolution are
desirable to trace ionospheric variations on shorter timescales and
smaller spatial scales.  For European telescopes, {\tt ionFR} can also
use TEC maps from the Royal Observatory of Belgium (ROB\footnote{{\tt
    http://gnss.be/Atmospheric\_Maps/ionospheric\_maps.php}}), which 
are derived from GPS data from a permanent European network.  These
TEC maps are more finely gridded than those from CODE
($\Delta_{\mathrm{lon.}}$ = 0.5$^\circ$, $\Delta_{\mathrm{lat.}}$ =
0.5$^\circ$) and have 15-minute time resolution (Fig.~\ref{euTECmap}).
They are also now publicly available via anonymous {\tt
  ftp}\footnote{{\tt ftp://gnss.oma.be/gnss/products/IONEX/}}, and 
are being produced since the beginning of 2012.  Comparisons of
{\tt ionFR}-modeled Faraday depth based on the CODE/ROB maps are discussed in
\S5.1 (see Figs.~\ref{LOFARobsmodelthree} and \ref{LOFARobsmodelint}).

\subsection{The geomagnetic field}\label{Ionogm}

The Earth's magnetic field is calculated using the eleventh generation
of the International Geomagnetic Reference Field
\citep[IGRF11;][]{2010GeoJI.183.1216F} released in December 2009.
The IGRF is derived by the International Association of Geomagnetism
and Aeronomy (IAGA) every five years and is available to
download\footnote{\tt{http://www.ngdc.noaa.gov/IAGA/vmod/igrf.html}}.
The IGRF11 is described as the negative gradient of a scalar
potential, B= $- \nabla V$, which is a finite series of spherical
harmonics.  {\tt ionFR} calls the IGRF11 to deliver the vector
components of the geomagnetic field at the IPP. These point towards
the north ($X$), east ($Y$), and radially towards the center of the
Earth ($Z$). The $XYZ$ coordinates are local (orthogonal) coordinates,
i.e. $X$ is not pointing in the direction of global north but to the
local (tangential) north on the sky. The total magnetic field along
the LOS at the IPP is then estimated as follows:

\begin{equation}
\mathrm{B}_{\rm LOS} = Z\mathrm{cos(\zeta^{\prime})} + Y
\mathrm{sin(\zeta^{\prime})}\mathrm{sin(\varphi^{\prime})} +
X\mathrm{sin(\zeta^{\prime})}\mathrm{cos(\varphi^{\prime})}.
\end{equation}

\subsection{Error propagation}\label{Ionoep}

The fractional uncertainties on $\mathrm{B}_{\rm LOS}$ compared with
their central values ($\sigma_{\mathrm{B}_{\rm LOS}} / \mathrm{B}_{\rm
  LOS}$) are much smaller than the fractional uncertainties in the
slant TEC ($\sigma_{\mathrm{TEC}_{\rm LOS}} / \mathrm{TEC}_{\rm
  LOS}$). Consequently, to determine the uncertainties on
$\phi_{\mathrm{ion}}$, only the RMS slant TEC values are used.  This
results in uncertainties of 0.1 -- 0.3 rad m$^{-2}$ in
$\phi_{\mathrm{ion}}$ using the CODE global TEC maps.  Assuming RMS
values of 0.5 TECU for the ROB European TEC maps results in smaller
uncertainties of 0.03 -- 0.06 rad m$^{-2}$.

\section{Ionospheric RM variation}

Here we present some examples of {\tt ionFR}-modeled ionospheric
Faraday depths using TEC data from CODE. We illustrate both the
$\phi_{\mathrm{ion}}$ variation within a day (for two epochs with
differing levels of solar activity) as well as the longer-term
variations with season and solar activity.

The modeled ionospheric Faraday rotations have been produced for
the geographic coordinates of the LOFAR core and the LOS towards the
supernova remnant Cassiopeia A (Cas A; RA =
23$^{\mathrm{h}}$23$^{\mathrm{m}}$27.9$^{\mathrm{s}}$, DEC =
+58$^{\circ}$48$^{\prime}$42.4$^{\prime\prime}$). Cas A was selected
because it is circumpolar as viewed from LOFAR (its minimum elevation
is $\sim 22^{\circ}$).  Therefore, {\tt ionFR} can calculate the
variation of $\phi_{\mathrm{ion}}$ for an entire day.

\begin{figure*}
\centering 

\includegraphics[trim = 20mm 0mm 25mm 5mm,clip=True,width=18cm]{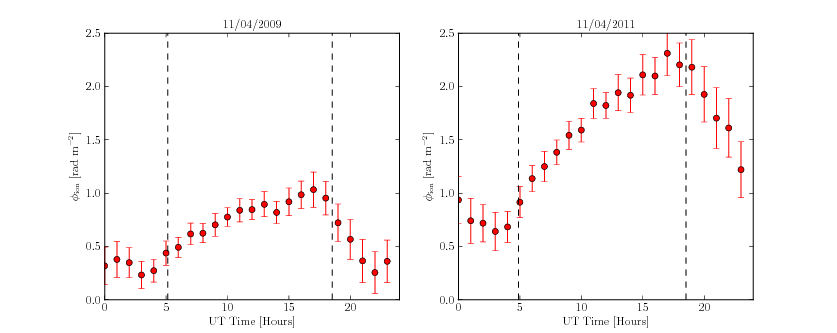} 

\caption{Prediction of the ionospheric Faraday rotation for two
  different epochs, as viewed from the LOFAR core along the LOS to Cas
  A. {\it Left}: Prediction when solar activity is near its
  minimum. {\it Right}: Prediction when the solar activity rises
  towards a new maximum, expected around May 2013. The dashed vertical
  lines mark the times of sunrise and sunset.}

\label{ionPredict}
\end{figure*} 

Figure~\ref{ionPredict} shows the variation in the modeled 
$\phi_{\mathrm{ion}}$ for two separate days: April 11th, 2009, close
to the most recent minimum in solar activity, and April 11th, 2011, by
which time solar activity had increase significantly.  The next solar
maximum is expected to be reached around mid 2013.  These plots show
the daily variation of $\phi_{\mathrm{ion}}$, increasing around
sunrise and decreasing around sunset. This is expected due to the
increase in the density of free electrons in the ionosphere from solar
irradiation during the day.  During night hours these free electrons
begin to recombine with free ions.  The degree of variation in
$\phi_{\mathrm{ion}}$ is noticeably different on the two days. Near
solar minimum (Fig.~\ref{ionPredict}, left), the daily peak
$\phi_{\mathrm{ion}}$ value is $\sim 1$ rad m$^{-2}$, while for the
prediction two years later (Fig.~\ref{ionPredict}, right), in which
solar activity has increased, the level sometimes exceeds 2 rad
m$^{-2}$.  Note that these time-variable contributions are much larger
than the formal uncertainties that are achievable through RM-synthesis
techniques applied to low-frequency radio data (see \S5).

\begin{figure*}

\includegraphics[trim = 13mm 5mm 13mm 10mm ,clip=True,width=9cm]{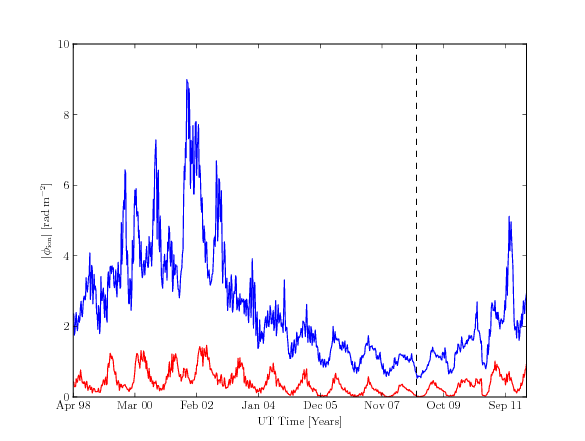}
\includegraphics[trim = 13mm 5mm 13mm 10mm ,clip=True,width=9cm]{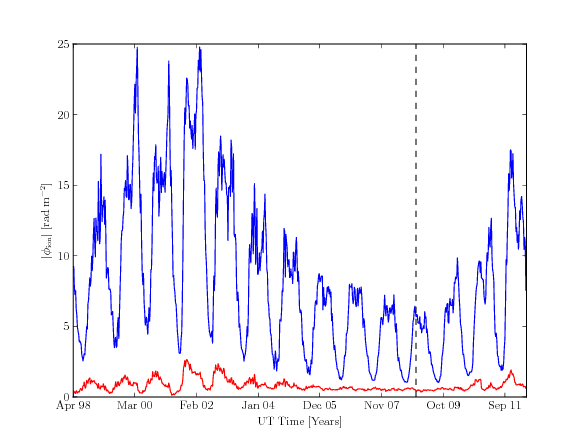}

\caption{ Weekly averages of the maximum and minimum (blue and red
  lines, respectively) absolute ionospheric Faraday depth
  $|\phi_{\mathrm{ion}}|$ from April 1998 -- 2012, as modeled by {\tt
    ionFR}.  The absolute value is shown because the ionospheric
  Faraday depth is positive or negative if observing from the northern
  or southern hemisphere, respectively.  The dashed vertical lines
  represent the end of Solar Cycle 23 (December 2008) and beginning of
  the current Solar Cycle 24.  {\it Left}: towards CasA, as viewed
  from LOFAR. {\it Right}: towards Eta Carinae, as viewed from an
  average of the SKA core sites in Western Australia and South
  Africa.}

\label{ionSolarCyc}
\end{figure*}

Figure~\ref{ionSolarCyc} (left) shows weekly averages of the daily
maximum and minimum modeled $|\phi_{\mathrm{ion}}|$ since April
1998, along the LOS of Cas A as seen from the LOFAR core. According
to the Space Weather Prediction Center (SWPC), Solar Cycle 23 started
in May 1996 and ended in December 2008.  {\tt ionFR} was run for each
day in Solar Cycles 23 and 24, until April 2012. The code produced
daily files, each containing 24 values of $\phi_{\mathrm{ion}}$. From
each file the daily maximum and minimum $|\phi_{\mathrm{ion}}|$ values 
were obtained.  These values were averaged to give a representative
maximum and minimum $|\phi_{\mathrm{ion}}|$ for every week since May
1996. Due to the incompleteness of GIMs within the IONEX files for the
years 1996, 1997, and part of 1998, the ionospheric predictions are
only shown since April 1998.  The oscillation in the minimum
$|\phi_{\mathrm{ion}}|$ is a well-known seasonal effect; more
ionization is expected in summertime than in winter.  In contrast, the
maximum $|\phi_{\mathrm{ion}}|$ reveals that during the years of
greatest solar activity (as reported by SWPC) several rad
m$^{\mathrm{-2}}$ can be reached, as viewed from the LOFAR core. It is
also noted that when solar activity is at its highest, the maximum
$|\phi_{\mathrm{ion}}|$ no longer appears to be dominated by seasonal
variations.

It is evident from Fig.~\ref{TECmap} that the ionosphere above the two
sites chosen for the SKA are subject to the Equatorial Ionization
Anomaly (EIA).  These two regions of enhanced plasma density are
located approximately 15 degrees north and south of the magnetic dip
equator \citep{1} and are the result of the equatorial fountain effect
\citep{2}.  We note that the SKA will suffer far higher levels of
ionospheric Faraday rotation than at the LOFAR sites, as the southern
component of the EIA can pass directly above both locations.
Figure~\ref{ionSolarCyc} (right) illustrates this by showing a
representative history of $|\phi_{\mathrm{ion}}|$ for the two proposed
SKA core sites.  Predictions for the two sites were made towards the
same astronomical object and then averaged.  The LOS chosen to
generate the two predictions was towards Eta Carinae (RA =
10$^{\mathrm{h}}$45$^{\mathrm{m}}$03.6$^{\mathrm{s}}$, DEC =
-59$^{\circ}$41$^{\prime}$04.0$^{\prime\prime}$), which is circumpolar
as viewed from both proposed sites. As expected, we observe that
$|\phi_{\mathrm{ion}}|$ can be much higher than for LOFAR
(Fig.~\ref{ionSolarCyc}, left). The seasonal effect is clearly visible
in the maximum $|\phi_{\mathrm{ion}}|$ curve.  Additionally,
Figure~\ref{TECmap} shows that during 12 hours the TEC and hence
ionospheric RM were subject to large variations ($> 80$ TECU).  These
facts underline the vital importance of ionospheric calibration for
polarimetric studies with the SKA --- particularly during the day, but
even at night.

Also, as seen in Fig.~\ref{TECmap}, the EIA passes very 
 close to the site of the GMRT, located near
  Pune, India.  Hence, polarimetric observations with the GMRT can
  also benefit greatly from {\tt ionFR}, especially for achieving the
  full potential of its available bands below 300MHz.

\section{RM-synthesis}\label{sec:RMsyn}

RM, which quantifies the amount of Faraday rotation along the LOS, has
commonly been determined as the gradient of the polarization angle
($\chi$) as a function of wavelength squared ($\lambda^{\mathrm{2}}$);
see, e.g., \citet{1962Natur.195.1084C},
\citet{1994MNRAS.268..497R}, and \citet{2011ApJ...728...97V}.
These previous RM measurements assumed therefore a linear relationship
between $\chi$ and
$\lambda^{\mathrm{2}}$. \citet{2005A&A...441.1217B} further
developed an alternative method to measure Faraday rotation called
RM-synthesis, which was first proposed by
\citet{1966MNRAS.133...67B} and is now increasingly used for such measurements
\citep[e.g.,]{2009A&A...503..409H,2011A&A...525A.104P}.  The
benefits of using RM-synthesis are numerous. For instance, it
minimizes or eliminates any n$\pi$ ambiguity, as opposed to the
`gradient' method in which $\chi$ can rotate by 180$^{\circ}$ an
arbitrary number of n times between data points at each
$\lambda^{\mathrm 2}$. It also uses the polarization information
across the entire observing bandwidth simultaneously such that it is
not necessary to detect polarization angles at each $\lambda^{\mathrm
  2}$. Additionally, it does not assume that there is a single RM
towards the LOS, i.e. that the linear relationship between $\chi$ and
$\lambda^{\mathrm 2}$ holds.

Following \citet{2005A&A...441.1217B} we define the observed complex polarization vector $P$ ($Q$ + $iU$) as:
\begin{equation}
\label{eq:Pint}
P(\lambda^{\mathrm{2}}) = \int^{\mathrm{\infty}}_{\mathrm{-\infty}} F(\phi)\exp\left[2\mathit{i}\phi\lambda^{\mathrm{2}}\right] \mathrm{d}\phi,
\end{equation}
where $F(\phi)$ is the intrinsic complex polarized surface brightness per unit Faraday depth,
known as the Faraday dispersion function \citep[FDF, or Faraday spectrum;][]{1966MNRAS.133...67B}.
Hence, Eq.~\ref{eq:Pint} can be inverted to obtain the FDF from the observed complex polarization vector:
\begin{equation}
\label{eq:FDF}
F(\phi) = \int^{\mathrm{\infty}}_{\mathrm{-\infty}} P(\lambda^{\mathrm{2}})\exp\left[-2\mathit{i}\phi\lambda^{\mathrm{2}}\right] \mathrm{d}\lambda^{\mathrm{2}}.
\end{equation}

\begin{figure*}[htp]
  \centering
\resizebox{\hsize}{!}{
\includegraphics[trim = 9mm 5mm 9mm 6mm,clip=True,width=7cm]{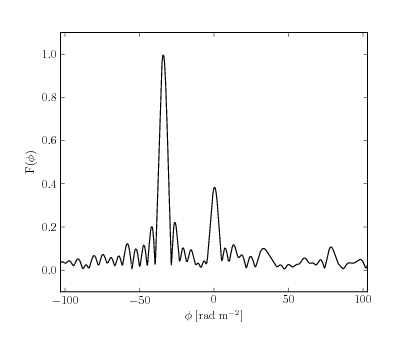}
\includegraphics[trim = 9mm 5mm 9mm 6mm,clip=True,width=7cm]{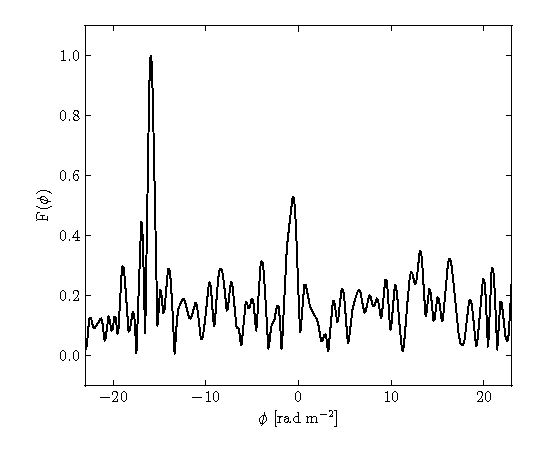}}

\caption{Examples of the absolute value of the normalized Faraday
  Dispersion Function (FDF), obtained from two LOFAR observations. The
  peaks at zero Faraday depth, $F(\phi\approx0)$, indicate the
  instrumental response.  {\it Left}: FDF obtained from a single
  3-minute HBA observation ($119-125$ MHz) of PSR B2217+47.  The peak
  at $F(\phi=-34.08~{\rm rad~m}^{-2})$ is the response due to the
  polarized flux of the pulsar.  The FWHM of the RMSF is 6.6 rad
  m$^{-2}$.  {\it Right}: FDF obtained from a single 3-minute LBA
  observation ($58-64$ MHz) of PSR B1919+21.  The polarized flux of
  the pulsar is responsible for the peak at $F(\phi=-15.94~{\rm
    rad~m}^{-2})$.  The FWHM of the RMSF is 0.84 rad m$^{-2}$.  }

\label{fig:FDF}
\end{figure*}

Polarization observations have a limited range in
$\lambda^{\mathrm{2}}$ and $\lambda^{\mathrm{2}}~\leq~0$ is not
possible.  Therefore, \citet{2005A&A...441.1217B} introduce a
window function that is non-zero for all observed
$\lambda^{\mathrm{2}}$ and zero otherwise.  In practice, the integral
in Eq.~\ref{eq:FDF} is performed as a discrete sum such that for each
discrete channel $i$ at $\lambda^{\mathrm{2}}_{i}$:
\begin{equation}
\label{eq:FDFsum}
F(\phi) \approx \left( \sum^{N}_{i=1} \omega_{i}\right)^{\mathrm{-1}} \sum^{N}_{i=1}  \omega_{i} P(\lambda_i^{\mathrm{2}})\exp\left[-2\mathit{i}\phi(\lambda_i^{\mathrm{2}}-\lambda_{\mathrm{0}}^{\mathrm{2}})\right],
\end{equation}
where $\omega_{i}$ is the weight of channel $i$ and
$\lambda_{\mathrm{0}}$ is the weighted average of the observed $\lambda^{\mathrm{2}}$.
Considering this inversion as the `coherent' addition of the observed polarization vectors
(a polarization vector per channel) 
for a range of Faraday depths, the vectors will constructively interfere within the bandwidth
for the Faraday depth of the observed source and will result in a peak at this $\phi$ in the FDF.
To illustrate this, example FDFs obtained from LOFAR High-Band Antenna (HBA) and Low-Band Antenna 
(LBA) observations (see \S5) are displayed in Fig.~\ref{fig:FDF}.  
Equation~\ref{eq:FDFsum} is used to determine the FDF for the observations described here.
The relationship between the input observed complex polarization vector and the output FDF 
is given by the rotation measure spread function (RMSF):
\begin{equation}
\label{eq:R}
R(\phi) \approx \left( \sum^{N}_{i=1} \omega_{i}\right)^{\mathrm{-1}} \sum^{N}_{i=1}  \omega_{i} \exp\left[-2\mathit{i}\phi(\lambda_i^{\mathrm{2}}-\lambda_{\mathrm{0}}^{\mathrm{2}})\right].
\end{equation}

RM-synthesis is analogous to performing aperture synthesis imaging with an
interferometer in the sense that both methods make use of Fourier transforms using
discrete sampling in $\lambda^{2}$ and spatial frequency coordinates ($u,v,w$), respectively 
\citep[e.g.,]{2009A&A...503..409H}.  Hence, the RMSF in Faraday space $\phi$ is 
analogous to the dirty beam in angular coordinates on the sky ($l,m$) \citep{2012A&A...540A..80B}.
As such, fewer gaps in the sampling of $\lambda^{2}$ reduce the side lobes in the RMSF 
and using larger bandwidths in $\lambda^{2}$ space increases the resolution in $\phi$ space.
The resolution in Faraday space can be quantified by the FWHM of the RMSF function, 
\begin{equation}
\label{eq:FWHM}
\mathrm{FWHM} \approx \frac{3.8}{\lambda^{\mathrm 2}_{\mathrm{max}} - \lambda^{\mathrm 2}_{\mathrm{min}}}~{\mathrm{rad~m^{\mathrm -2}}},
\end{equation}
where $\lambda_{\mathrm{min}}$ and $\lambda_{\mathrm{max}}$ are the 
shortest and longest observed wavelengths
and the constant 3.8 is from \citet{2009A&A...494..611S}.
Moreover, the uncertainty associated with
locating the peak Faraday depth in the FDF can be determined in the
same way as the uncertainty associated with locating the peak flux in an
aperture synthesis image \citep[see]{1999ASPC..180..301F}:
\begin{equation}
\label{eq:noiseFD}
\sigma_{\phi} = \frac{\mathrm{FWHM}}{2 \times \mathrm{S/N}},
\end{equation} 
where FWHM is the FWHM of the RMSF defined in Eq.~\ref{eq:FWHM} 
and S/N is the total polarized signal-to-noise ratio. 
Equation~\ref{eq:noiseFD} is used to determine the error in the Faraday depth measurements presented here.

\section{Model comparison with observational data}

\subsection{LOFAR pulsar data \label{lofar_psr}}

To measure Faraday depth variations in the ionosphere and to compare
them with those modeled by \texttt{ionFR}, four bright polarized
pulsars were observed using LOFAR \citep[see][for a description of
  LOFAR's pulsar observing modes]{2011A&A...530A..80S}.  One or more
pulsars were observed on four separate epochs, including times when the 
ionosphere was expected to be particularly dynamic (around sunrise and
sunset).  See Table~\ref{tb:obs} for a summary of these four observing
campaigns.
 
The first campaign used the coherently combined LOFAR
`Superterp'\footnote{The 330-meter-wide inner core of the array, which
  hosts 6 stations.}  to observe PSR B0834+06 in $7 \times 10$-min
integrations spaced every 50 min.  These started 1.8 hr before sunset
(18:25 UT) and continued until more than 2 hr after astronomical
twilight (20:40 UT).

The second Superterp campaign observed PSR B0834+06 in $20 \times
3$-min integrations spaced every 7 min.  These started after
astronomical twilight (04:11 UT) and continued until 1.5 hours after
sunrise (06:08 UT).

The third Superterp campaign observed PSRs B1642$-$03, B1919+21 and
B2217+47 by cycling consecutively through the three pulsars so that
each was observed for $12 \times 3$-min integrations spaced every 20
min.  This enabled quasi-simultaneous measurements of the ionospheric
Faraday depth variations towards three widely separated LOSs.  These
observations started before nautical twilight (04:16 UT) and ended two
hours after sunrise (05:29 UT).

The fourth campaign observed PSR B0834+06 using the LOFAR Superterp
stations and two international stations located near Nan\c{c}ay,
France and near Onsala, Sweden.  The pulsar was quasi-simultaneously
observed by each station in $11 \times 3$-min integrations spaced by
17 min.  These were done during midday when the absolute TEC was
expected to be relatively high.  This enabled measurements of Faraday
depth variations from three locations separated by long geographical
baselines -- 594 km minimum and 1294 km maximum distance, respectively.

\begin{center}
\begin{table*}[ht]
{\scriptsize
\hfill{}
\begin{tabular}{@{}c c c c c c c c c c c c@{}}
\hline\hline\\[-0.8em]                 
No. & PSR &  Date & Obs. duration & Sun[rise,set] time & LOFAR stations & Time obs. & No. obs. & Freq & Elevation & LOFAR obs. IDs\\
 & (B-name) & (dd.mm.yyyy) & (hh:mm UT) & (hh:mm UT) & & (min) &  & (MHz) & (deg) & \\[1pt]
\hline\\[-0.8em]
1 & B0834+06 & 11.04.2011 & 16:40 -- 22:50 & 18:25 & CS00[2-7]HBA & 10 & 7 & 120--126 & 20 -- 45 & L25152 -- L25158 \\      
2 & B0834+06 & 20.10.2011 & 04:20 -- 07:33 & 06:08 & CS00[2-7]HBA & 3 & 20 & 129--140 & 37 -- 44 & L32350 -- L32369 \\
3 & B1642$-$03 & 23.03.2012 & 04:08 -- 07:40 & 05:29 &  CS00[2-7]HBA & 3 & 12 & 119--125 & 16 -- 34 & L53966 -- L53977\\
3 & B1919+21 & 23.03.2012 & 04:04 -- 07:36 & 05:29 & CS00[2-7]LBA & 3 & 12 & 58--64 & 45 -- 59 & L53942 -- L53953\\
3 & B2217+47 & 23.03.2012 & 04:12 -- 07:44 & 05:29 & CS00[2-7]HBA & 3 & 12 & 119--125 & 37 -- 72 & L53990 -- L54001\\
4 & B0834+06 & 10.07.2012 & 11:20 -- 14:43 & N/A & CS00[2,3,5-7]HBA & 3 & 11 & 119--129 & 38 -- 44 & L61473 -- L61483\\
4 & B0834+06 & 10.07.2012 & 11:25 -- 14:48 & N/A & FR606HBA & 3 & 11 & 119--129 & 42 -- 49 & L61532 -- L61542\\
4 & B0834+06 & 10.07.2012 & 11:30 -- 14:53 & N/A & SE607HBA & 3 & 11 & 119--129 & 32 -- 39 & L61520 -- L61530\\[1pt]
\hline
\end{tabular}}
\hfill{}

\caption{Summary of the four LOFAR pulsar observing campaigns.
  Columns 1--11 show the number of the observing campaign, pulsar B
  names, date, duration, time of sunset or sunrise during the
  observations, LOFAR station(s) (CS00[2-7] indicate the LOFAR
  `Superterp' stations combined in tied-array mode, HBA and LBA
  indicate High-Band Antenna stations and Low-Band Antenna stations,
  respectively), individual observation integration times, total
  number of observations, frequency range, elevation range, and LOFAR
  observation identification numbers (obs. IDs).  }

\label{tb:obs}
\end{table*}
\end{center}

In all cases, data were written as complex values for the two
orthogonal linear polarizations.  The data were recorded using the 200
MHz clock mode, which provides multiple 195.3125 kHz subbands that are
further channelized by an online poly-phase filter to 12.2 kHz
channels with a time resolution of 81.92 $\upmu$s.  Due to limitations
on the data rate at the time of observation, $6-11$ MHz of bandwidth
were recorded.  In comparison, 80 MHz of
bandwidth can now be recorded by LOFAR in this mode.  Nonetheless,
  given the low central observing frequencies ($\sim 125$ MHz) the
  recorded bandwidths were still more than adequate to achieve precise
  RM measurements.  The complex values were converted to 8-bit samples
  offline and then coherently dedispersed and folded using the {\tt
    dspsr} program \citep{2011PASA...28....1V}.  Radio frequency
  interference (RFI) was removed using the {\tt pazi} program of
  PSRCHIVE \citep{2004PASA...21..302H}.

The reduced data were analyzed using an RM-synthesis program in order
to determine a precise Faraday depth for each individual observation.
For each data set, the Stokes parameters ($IQUV$) and associated
uncertainties for each frequency channel were output for the pulsed
section of the pulsar profile using the PSRCHIVE program {\tt rmfit}.
The frequency, $Q$ and $U$ information were used as the input to the
RM-synthesis program, which calculated the FDF as a discrete sum for
Faraday depths $-$50 $\leq$ $\phi$ $\leq$ 50 rad m$^{-2}$ in steps of
0.001 rad m$^{-2}$ using Eq.~\ref{eq:FDFsum} (see Fig.~\ref{fig:FDF}
for two examples of these).  The peak associated with the instrumental
DC signal at $\sim$0 rad m$^{-2}$ Faraday depth \citep[see,
  e.g.,]{2011MNRAS.418..516G} was subtracted from the FDF before
determining the peak associated with the Faraday depth towards each
pulsar LOS.  This had no effect on the Faraday peaks of the pulsars
since the known RM values \citep[ATNF pulsar
  catalog;][]{2005yCat.7245....0M} are $>$ $2\times$FWHM of the RMSF,
see Eq.~\ref{eq:FWHM}.  The Faraday depth at which the peak in the FDF
occurred was assumed to be the measured Faraday depth of the ISM
and ionosphere towards the pulsar,
$\phi_{\mathrm{ISM}}+\phi_{\mathrm{ion}}$\footnote{This assumes no
  Faraday rotation in the pulsar magnetosphere itself; see
  \citealp{2009MNRAS.396.1559N} and \citealp{2011MNRAS.417.1183W} for
  a discussion on possible Faraday rotation within pulsar
  magnetospheres.}.  For each observing campaign, we estimated the
instrumental error by measuring the Faraday depth of the instrumental
polarization peak around 0 rad m$^{-2}$ in the FDF of each
observation, weighted by the S/N, and taking the 0 rad m$^{-2}$
$\pm$1$\sigma$ limits (Table~\ref{tb:1}).  This demonstrated that
larger bandwidth observations with higher S/Ns also tended to reduce
the scatter in instrumental Faraday depth around 0 rad m$^{-2}$.  The
total error on the Faraday depth was taken to be the formal error from
Eq.~\ref{eq:noiseFD} added in quadrature with the instrumental error.
The linear polarization had S/N $> 30$ for all observations; this is
well above the threshold necessary for reliable Faraday depth
measurements \citep[see][and references
  therein]{2012arXiv1203.2706M}. While we have shown that LOFAR 
provides reliable Faraday depths, we note that absolute polarization calibration 
(e.g. to determine absolute polarization angles) has not yet been applied to the data.

\begin{figure*}[!htb]
\resizebox{\hsize}{!}{\includegraphics{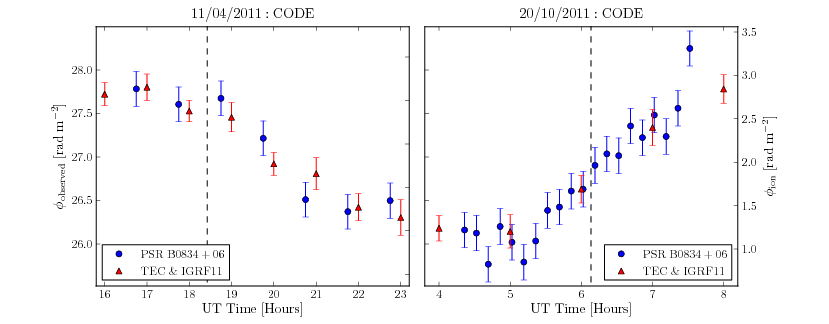}}

\caption{Observed Faraday depths, $\phi_{\mathrm{observed}}$, and
  \texttt{ionFR}-modeled ionospheric Faraday depths,
  $\phi_{\mathrm{ion}}$, towards PSR B0834+06 as a function of time
  (observations: blue circles, left axis labels; model: red triangles,
  right axis labels).  Only CODE TEC data was available for these
  predictions.  {\it Left panel:} seven LOFAR Superterp HBA
  observations during sunset.  {\it Right panel:} twenty LOFAR
  Superterp HBA observations during sunrise.  The vertical dashed
  lines indicate the time of sunset and sunrise, respectively. The
  offsets between $\phi_{\mathrm{observed}}$ and $\phi_{\mathrm{ion}}$
  for the sunset and sunrise campaigns are 25.15 $\pm$ 0.18 and 24.94
  $\pm$ 0.24 rad m$^{\mathrm{-2}}$, respectively.  }

\label{LOFARobsmodel}
\end{figure*}

\begin{figure*}[htb]
\centering
\resizebox{\hsize}{!}{
  \subfigure[]{\includegraphics[trim = 2mm 15mm 2mm 10mm,clip=True,width=9cm]{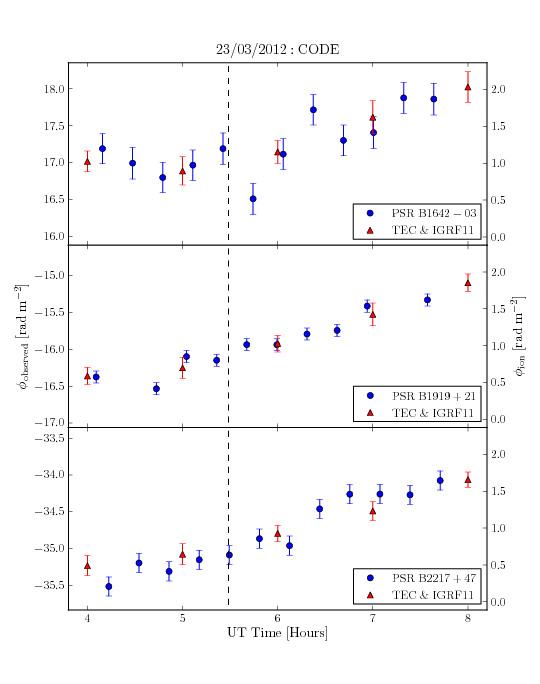}}\hfil
  \subfigure[]{\includegraphics[trim = 2mm 15mm 2mm 10mm,clip=True,width=9cm]{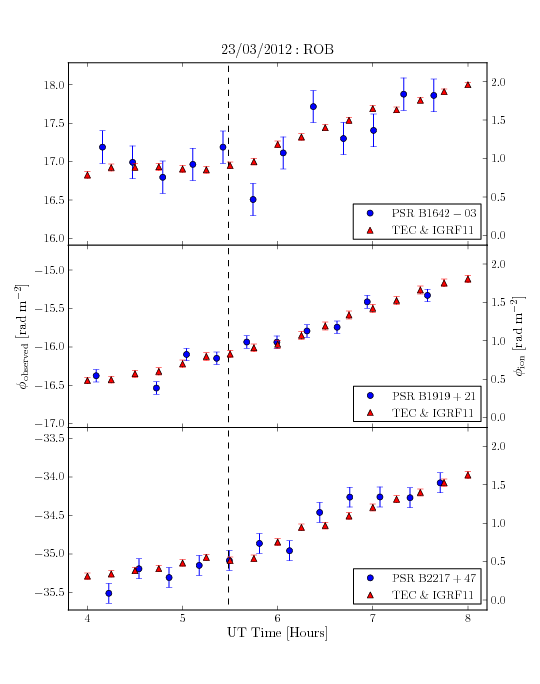}}}

\caption{ Observed Faraday depths, $\phi_{\mathrm{observed}}$, and
  \texttt{ionFR}-modeled ionospheric Faraday depths,
  $\phi_{\mathrm{ion}}$, towards three pulsars as a function of time
  during sunrise (observations: blue circles, left axis labels; model:
  red triangles, right axis labels).  {\it Upper panels:} twelve LOFAR
  Superterp HBA observations of PSR B1642$-$03.  {\it Middle panels:}
  twelve LOFAR Superterp LBA observations of PSR B1919+21.  {\it Lower
    panels:} twelve LOFAR Superterp HBA observations of PSR B2217+47.
  The vertical dashed lines indicate the time of sunrise.
  $\mathrm{(a)}$ shows the {\tt ionFR} model using CODE TEC data and
  IGRF11.  $\mathrm{(b)}$ shows the {\tt ionFR} model using ROB TEC
  data and IGRF11.  }

\label{LOFARobsmodelthree}
\end{figure*}

\begin{figure*}[htb]
\centering
\resizebox{\hsize}{!}{
  \subfigure[]{\includegraphics[trim = 2mm 15mm 2mm 10mm,clip=True,width=9cm]{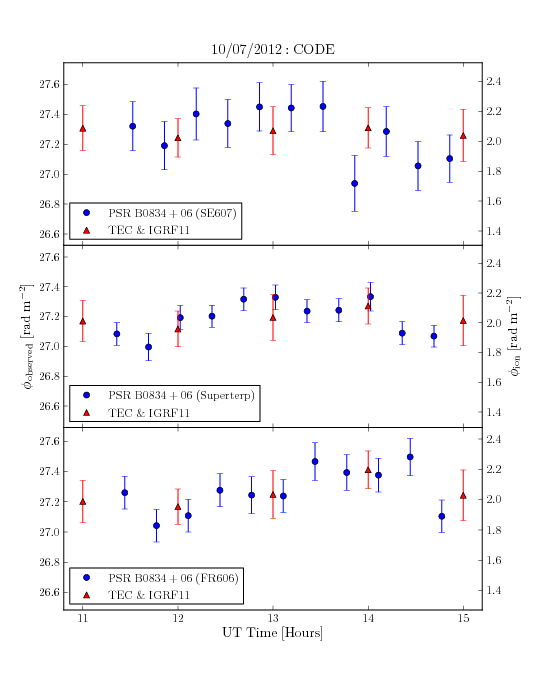}}\hfil
  \subfigure[]{\includegraphics[trim = 2mm 15mm 2mm 10mm,clip=True,width=9cm]{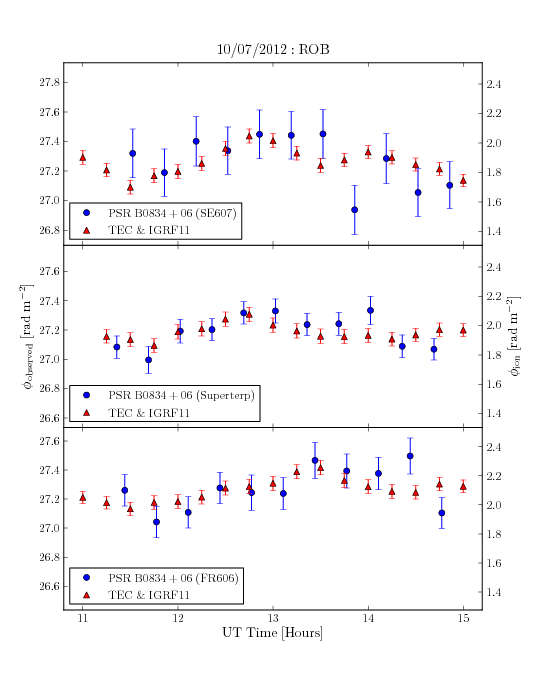}}}

  \caption{ Observed Faraday depths, $\phi_{\mathrm{observed}}$, and
    \texttt{ionFR}-modeled ionospheric Faraday depths,
    $\phi_{\mathrm{ion}}$, towards B0834+06 as a function of time
    during midday (observations: blue circles, left axis labels;
    model: red triangles, right axis labels).  {\it Upper panels:}
    eleven LOFAR HBA observations using the international station near
    Onsala, Sweden.  {\it Middle panels:} eleven LOFAR Superterp HBA
    observations.  {\it Lower panels:} eleven LOFAR HBA observations
    using the international station near Nan\c{c}ay, France.
    $\mathrm{(a)}$ shows the {\tt ionFR} model using CODE TEC data and
    IGRF11.  $\mathrm{(b)}$ shows the {\tt ionFR} model using ROB TEC
    data and IGRF11.  }

\label{LOFARobsmodelint}
\end{figure*}

The observed and \texttt{ionFR}-modeled Faraday depths as a function
of time for the four LOFAR observing campaigns are plotted in
Figs.~\ref{LOFARobsmodel}, \ref{LOFARobsmodelthree} and
\ref{LOFARobsmodelint}.  In general, the modeled and observed Faraday
depth variations agree very well.  However, there are a few instances
  where the observed and modeled values differ by more than
  1$\sigma$.  The measured RMs could still
  be affected by interference in some cases and it is also quite possible that there
  are unmodeled ionospheric variations on short timescales (see also
  \S\ref{disc}).  TEC data from CODE was used for the B0834+06 sunset
and sunrise campaigns.  CODE and ROB TEC data were both available for
the observations which took place in 2012 and are compared in
Figs.~\ref{LOFARobsmodelthree} and \ref{LOFARobsmodelint}.  In each
case, there is a clear trend in the Faraday depth as a function of
time.  At sunrise and sunset there is a particularly distinct
variation in Faraday depth of approximately 2 rad m$^{\mathrm{-2}}$,
as expected (cf. Fig.~\ref{ionPredict}).  Even during midday,
Fig.~\ref{LOFARobsmodelint}, when the absolute TEC is expected to be
high but relatively constant
\citep[e.g.,][Fig.~14]{2007ApJS..172..686K}, smaller variations in
Faraday depth are still evident and well modeled in general.

Figs.~\ref{LOFARobsmodelthree} and \ref{LOFARobsmodelint} compare the
modeled Faraday depths using CODE and ROB TEC data.  In both cases, the
observations and model show good agreement, although it is evident
that there are variations on timescales less than one hour which are
not resolved in the model using CODE data.  The model using ROB TEC
data more accurately fits the observations.  The finer griding and
smaller RMS available with the ROB TEC data also allows smaller
fluctuations in Faraday depth to be resolved.  For the observations of
B0834+06 using the Superterp and international stations over midday,
Fig.~\ref{LOFARobsmodelint}, this is especially significant, where the
variations in the data appear on shorter timescales and in Faraday
depths within the error bars of the model using CODE data.
The model output of {\tt ionFR} shows good agreement with the measured
Faraday depths for all of the LOFAR observing campaigns depicted in
Figs.~\ref{LOFARobsmodel}, \ref{LOFARobsmodelthree} and
\ref{LOFARobsmodelint}. Therefore, this model was used to subtract 
the contribution of the ionospheric Faraday depth from the measurements 
in order to determine the Faraday depth of the ISM, 
$\phi_{\mathrm{ISM}}$, in the direction of these pulsars.

To determine $\phi_{\mathrm{ISM}}$ for each pulsar, the constant
offset which yielded the minimum weighted chi-squared value between
the pulsar Faraday depth measurements and the ionospheric modeled
Faraday depths from \texttt{ionFR} was used (see Table~\ref{tb:1}).
Using the CODE data, the reduced chi-squared values range from 0.4 --
1.0, whereas with the ROB data the reduced chi-squared values range
from 0.9 -- 1.3.  The residual differences between the observations
and model may be due to small scale variations in the ionosphere which
affect the observations but which are not resolved due to the time
resolution of the TEC data.  The standard deviations of
$\phi_{\mathrm{observed}}-\phi_{\mathrm{ion}}$ divided by the square
root of the number of measurements in each observational campaign
range from 0.03 -- 0.09 rad m$^{-2}$ for the CODE output and from 0.02
-- 0.07 rad m$^{-2}$ for the ROB output.  This indicates that both fit
the measurements well, although the ROB data gives reduced chi-squared
values closer to 1 and smaller errors due to the smaller uncertainties
compared with the CODE maps.  Also, both sources of TEC maps give
consistent results, where all $\phi_{\mathrm{ISM}}$ values obtained
using CODE and ROB data agree within 1-$\sigma$ for the same observing
campaign and LOS.  The Faraday depth values for B0834+06 obtained from
the three observational campaigns over six months apart also agree at
or below the 2-$\sigma$ level for both CODE and ROB data.  The
weighted mean for all CODE and ROB data available for B0834+06 gives
$\phi_{\mathrm{ISM}} = 25.12 \pm 0.07$ rad m$^{-2}$ (5 LOSs) and
$\phi_{\mathrm{ISM}} = 25.26 \pm 0.05$ rad m$^{-2}$ (3 LOSs),
respectively. Although there are two more data sets with CODE data
available, the errors assumed for the ROB data are smaller.  It is
worth noting that these are among the most precise pulsar-derived
$\phi_{\mathrm{ISM}}$ measurements ever obtained \citep[see ATNF
  catalog]{2005yCat.7245....0M}, and improve significantly on the
precision of previous RM measurements for these pulsars.

In order to determine the Faraday depth of B0834+06 using an
independent instrument, on April 22nd, 2012 at 16:50 UT we performed a
20-minute observation of PSR B0834+06 using the WSRT with the PuMaII pulsar backend
\citep{2008PASP..120..191K} from 310 to 390 MHz.  We recorded
baseband data, which were folded, dedispersed and
subsequently analyzed using the same RM-synthesis method described
above, yielding $\phi_{\rm observed} = 28.2 \pm 1.8$ rad m$^{-2}$.
The {\tt ionFR} code predicts $\phi_{\mathrm{ion}} = 2.75 \pm 0.15$
rad m$^{-2}$, using CODE maps, for the given time and LOS of this observation, resulting
in a corrected value of $\phi_{\mathrm{ISM}}^{WSRT} = 25.4 \pm 1.8$
rad m$^{-2}$. Using ROB maps we calculated a 
corrected value of $\phi_{\mathrm{ISM}}^{WSRT} = 25.1 \pm 1.5$
rad m$^{-2}$.
These values are in excellent agreement with the more precise
value derived from the LOFAR observations. The precision of the WSRT
measurement is lower in part because of the higher observing
frequency, see Eqs.~\ref{eq:FWHM} and \ref{eq:noiseFD}.  This
demonstrates the power of low-frequency observations for the purpose
of determining accurate $\phi_{\mathrm{ISM}}$.

Our measured Faraday depths are consistent with previously published
measurements for these pulsars, but are significantly more precise.
The values of $\phi_{\mathrm{ISM}}$ obtained from the LOFAR
observations of B0834+06 agree to within 2.5~$\sigma$ of the catalog
value for this pulsar, RM$_{\mathrm{psrcat}}$ = 23.6 $\pm$ 0.7 rad
m$^{-2}$.  It is unclear, but likely, that the catalog value was
calibrated for ionospheric Faraday rotation
\citep[see][\S2]{1987MNRAS.224.1073H}.  That paper also states that
the ionosphere contributes 1$-$8 rad m$^{-2}$ to other observations,
using either a geostationary satellite or an ionosonde, and that the
subtraction of this introduces uncertainties of approximately 1$-$2
rad m$^{-2}$.  The 2.5-$\sigma$ difference between the
$\phi_{\mathrm{ISM}}$ obtained using LOFAR and the ATNF catalog value
determined in 1987, plus the possibility that pulsar RMs may change on
multi-year timescales
\citep[e.g.,]{2004ApJS..150..317W} due
  to variations in the polarized pulsar emission and/or electron density
  changes along the LOS through the ISM, provided the motivation for
the recent independent comparison observation of PSR B0834+06 using
the WSRT.

The values of $\phi_{\mathrm{ISM}}$ obtained for the three pulsars
observed quasi-simultaneously are also very precise and are in
excellent agreement with those of the ATNF pulsar catalog.  Prior
measurements for B1642$-$03 and B1919+21, also calibrated for
ionospheric Faraday rotation using geostationary satellite and
ionosonde data, give RM$_{\mathrm{ISM}} = 15.8 \pm 0.3$ rad m$^{-2}$
and RM$_{\mathrm{ISM}} = -16.5 \pm 0.5$ rad m$^{-2}$, respectively
\citep{1987MNRAS.224.1073H}, and are also in good agreement within
1~$\sigma$ of the values obtained in this work (ROB-subtracted values
$\phi_{\mathrm{ISM}} = 16.04 \pm 0.18$ rad m$^{-2}$ and
$\phi_{\mathrm{ISM}} = -16.92 \pm 0.07$ rad m$^{-2}$).  A prior
measurement for B2217+47, RM$_{\mathrm{ISM}} = -35.3 \pm 1.8$ rad
m$^{-2}$, also calibrated for ionospheric Faraday rotation using a
geostationary satellite \citep{1972ApJ...172...43M}, is also in
good agreement within 1~$\sigma$ of the value derived here
(ROB-subtracted value $\phi_{\mathrm{ISM}} = -35.60 \pm 0.11$ rad
m$^{-2}$).

Together these observations provide a convincing verification of the
accuracy of the \texttt{ionFR} model and demonstrate the ability to
derive Faraday depths resulting solely from the ISM by robustly
removing the time and direction-dependent Faraday depth introduced by
the ionosphere, even during some of the most turbulent periods
expected in daily ionospheric variations; see \S6 for further
discussion.

\begin{center}
\begin{table*}[ht]
{\footnotesize
\hfill{}
\begin{tabular}{@{}c c c c c c c c c c c@{}}
\hline\hline\\[-0.8em]                 
PSR & Date  & Telescope  & Station(s) & FWHM$_{RMSF}$ & $\sigma_{\mathrm{inst}}$ & $\phi_{\mathrm{ISM}}^{\mathrm{CODE}}$ & 
$\chi^2_{\mathrm {red}}$ & $\phi_{\mathrm{ISM}}^{\mathrm{ROB}}$ & $\chi^2_{\rm red}$ & RM$_{\mathrm{ISM}}^{\mathrm{psrcat}}$ \\
  &  [dd.mm.yy] &  &  & [rad~m$^{-2}$] & [rad~m$^{-2}$]& [rad~m$^{-2}$] & CODE & [rad~m$^{-2}$] & ROB & [rad~m$^{-2}$]\\[1pt]
\hline\\[-0.8em]
    B0834+06 & 11.04.11 & LOFAR & Superterp & 6.2 & 0.20 & 25.15 $\pm$0.18 & 0.52 & N/A & N/A & 23.6 $\pm$ 0.7\\      
    B0834+06 & 20.10.11 & LOFAR & Superterp & 9.4 & 0.20 & 24.94 $\pm$ 0.24 & 0.64 & N/A & N/A & 23.6 $\pm$ 0.7\\
    B1642$-$03 & 23.03.12 & LOFAR & Superterp & 6.6 & 0.20 & 15.98 $\pm$ 0.23 & 0.8 &  16.04 $\pm$ 0.18 & 1.3 & 15.8 $\pm$ 0.3\\
    B1919+21 & 23.03.12 & LOFAR & Superterp & 0.8 & 0.08& -16.95 $\pm$ 0.12 & 0.5 & -16.92 $\pm$ 0.07 & 1.3 & -16.5 $\pm$ 0.5 \\
    B2217+47 & 23.03.12 &  LOFAR & Superterp & 6.6 & 0.13 & -35.72 $\pm$ 0.15 & 1.0 & -35.60 $\pm$ 0.11 & 1.1 & -35.3 $\pm$ 1.8\\
    B0834+06 & 22.04.12 &  WSRT & N/A & 12.0 & 1.2 & 25.4 $\pm$ 1.8 & N/A & 25.1 $\pm$ 1.5 & N/A & 23.6 $\pm$ 0.7\\
    B0834+06 & 10.07.12 & LOFAR & Superterp & 4.2 & 0.07 & 25.16 $\pm$ 0.13 & 0.5 & 25.23 $\pm$ 0.08 & 0.9 & 23.6 $\pm$ 0.7\\
    B0834+06 & 10.07.12 & LOFAR & FR606 & 4.2 & 0.10 & 25.21 $\pm$ 0.15 & 0.4 & 25.16 $\pm$ 0.10 & 1.2 & 23.6 $\pm$ 0.7\\
    B0834+06 & 10.07.12 & LOFAR & SE607 & 4.2 & 0.15 & 25.22 $\pm$ 0.18 & 0.6 & 25.39 $\pm$ 0.14 & 1.0 & 23.6 $\pm$ 0.7\\[1pt]
\hline
\end{tabular}}
\hfill{}

\caption{Summary of results from the four LOFAR pulsar observing
  campaigns and WSRT observation.  Columns 1--11 show the pulsar
  observed, date (see Table~\ref{tb:obs} for specific times), the
  telescope, the LOFAR stations (if applicable, see Table~\ref{tb:obs}
  for specific station names), the FWHM of the RMSF from RM-synthesis,
  the error introduced by instrumental effects
  $\sigma_{\mathrm{inst}}$, the Faraday depth of the ISM towards the
  pulsar $\phi_{\mathrm{ISM}}$ as determined using the CODE TEC maps,
  the reduced chi-squared value $\chi^2_{\mathrm {red}}$ for
  comparison between the ionospheric Faraday depths from
  \texttt{ionFR} using CODE TEC data and the observed Faraday depths,
  the Faraday depth of the ISM towards the pulsar
  $\phi_{\mathrm{ISM}}$ as determined using the ROB TEC maps, the
  reduced chi-squared value for comparison between the ionospheric
  Faraday depth produced from \texttt{ionFR} using CODE TEC data and
  the observed Faraday depths.  The final column gives the catalog RM
  value from psrcat \citep{2005yCat.7245....0M} for comparison.
  Note that ROB TEC maps were not available for the observations of
  B0834+06 taken in 2011.  Since a single observation was obtained
  with WSRT, the reduced chi-squared value is not applicable because
  the Faraday depth of the ISM was determined by subtracting the
  Faraday depth of the ionosphere from the observed Faraday depth.}

\label{tb:1}
\end{table*}
\end{center}

\subsection{WSRT imaging data}

Here we compare the {\tt ionFR}-modeled Faraday depths with archival
118.2-MHz WSRT observations of PSR B1937+21.

\begin{figure}
\resizebox{\hsize}{!}{\includegraphics{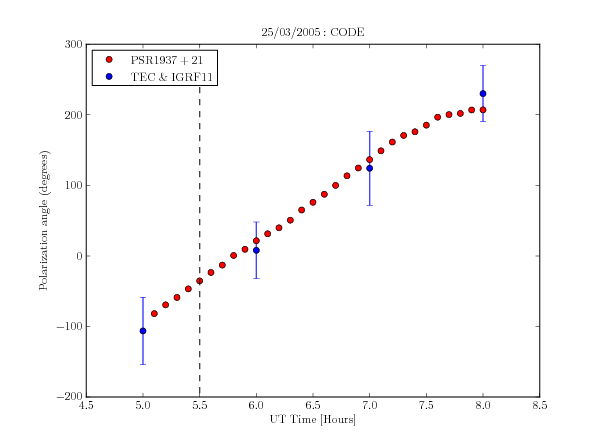}} 

\caption{PSR B1937+21's observed polarization angle compared with the
  predicted amount of rotation due to the ionosphere (red and blue
  points, respectively).  The errors on the polarization angles for
  the WSRT data are less than 2 degrees and are not visible in the
  plot. The vertical dashed line marks the time of sunrise.}

\label{fig:wsrt}
\end{figure}

PSR B1937+21 was observed with the WSRT Low Frequency Front Ends on 25
March 2005 for 3 hours following sunrise to monitor ionospheric
Faraday rotation. The observed polarization angle at 118.2 MHz is
shown in Fig.~\ref{fig:wsrt} and it is seen to increase rapidly and
smoothly.  The polarization angle presented refers only to a 4 MHz
band, from 116.05 -- 120.34 MHz. The calibrated data were split into 30
time slots, of 6 minutes each, in order to produce a pair of Stokes
$Q$ and $U$ maps for each slot.  From the polarization maps we then
calculated the polarization angle of PSR B1937+21.

Fig.~\ref{fig:wsrt} also shows the predicted rotation due to the
ionosphere. This was calculated by modeling the $\phi_{\mathrm{ion}}$
using CODE TEC data (no ROB data is available prior to 2012), and then
converting the ionospheric Faraday depths into polarization angles at
118.2 MHz.  After removing a constant offset, the modeled Faraday depths and
observation match very well.  This comparison provides further
verification of {\tt ionFR}-modeled Faraday depths and also shows that the
ionospheric correction can be done for interferometric imaging data.

As this example shows, low-frequency observations lasting a few tens
of minutes or more suffer depolarization from variable ionospheric
Faraday rotation; in this case, the polarization angle has made almost
a full turn within 3hrs.  The time-variable ionospheric Faraday depth
will lead to misalignment between the polarization vectors, which
therefore results in significant signal loss.  Even an instantaneous
measurement gives a Faraday depth that is systematically biased by
$\pm0.25 - 2.5$ rad m$^{-2}$ or more, which can be a large fraction of
the total measured Faraday depth for nearby sources ($d < 2$ kpc),
greatly impeding our ability to use these as probes of the local
interstellar magnetic field.

\section{Discussion}\label{disc}

The Faraday depths of the ISM towards PSRs B0834+06, B1642$-$03,
B1919+21 and B2217+47 presented here (Table~\ref{tb:1}) are some of
the first determined using the RM-synthesis technique and are among
the most precise Faraday depth measurements ever made towards a
pulsar.  In the ATNF catalog, there are only a dozen measurements with
absolute precision $\leq 0.2$ rad m$^{-2}$, determined for some of the
brightest known pulsars.  The fact that the Faraday depths obtained
here are also in excellent agreement with the catalog and/or
alternatively determined values also demonstrates the robustness of
{\tt ionFR} for subtracting the $\phi_{\mathrm{ion}}$ contribution and
hence for allowing us to reap the benefits of low-frequency
measurements.

Using multi-beaming capabilities \citep{2011A&A...530A..80S}, LOFAR
will provide high-precision Faraday depth measurements for hundreds of
nearby pulsars, which can be used to probe the interstellar magnetic
field of the Galaxy (e.g. Sobey et al. 2013, in preparation).  Since
the observations presented here used just a tenth of the now-available
LOFAR bandwidth, the prospects for even higher-precision Faraday depth
measurements are excellent, especially using LOFAR LBA observations
($10-90$ MHz).  However, with the TEC data available from ROB these
will be limited to precisions of approximately 0.05 rad m$^{-2}$
because of the systematic uncertainty provided by the RMS VTEC maps.
In other words, more sophisticated calibration techniques will need to
be devised in order to reap the {\it full} benefit of LOFAR's low
observing frequencies and large fractional bandwidth.  For instance,
it is possible to measure time-dependent, differential TEC and Faraday
rotation between LOFAR stations using visibility data from
observations of bright (even unpolarised) calibrators. These data can
be combined with magnetic field models in order to predict the
absolute Faraday depth, in addition to using the TEC data described
here for comparison. This is currently being investigated using LOFAR
long-baseline observations, but will be less suitable for more compact
array designs (e.g., LOFAR pulsar observations typically use only the
2-km core of the array).  Higher-precision Faraday depth measurements
are especially important for determining possible long-term variations
of the Faraday depth of pulsars due to fluctuations in their
magnetosphere or the ISM \citep[e.g.,]{2004ApJS..150..317W},
particularly as some pulsars have large relative velocities.

Calibrating for $\phi_{\mathrm{ion}}$ is also important for
higher-frequency observations.  Assuming a bandwidth of 1.21 -- 1.51
GHz, such as that of the multibeam receiver used at the 
100-m Effelsberg radio telescope \citep{2013MNRAS.429.1633B}
, the theoretical FWHM of the RMSF is 173 rad m$^{-2}$.
This yields an uncertainty in Faraday depth of 8.7 rad m$^{-2}$ given
a S/N of 10, which is comparable to the $\phi_{\mathrm{ion}}$ reached
during solar maximum as Fig.~\ref{ionSolarCyc} (left) shows.  Given
larger bandwidths, such as those planned for the SKA and its
pathfinders, this problem becomes worse and is relevant even during
night-time and solar minimum.  It is also clear that corrections for
the ionospheric Faraday depth are important for observations of
pulsars and extragalactic sources, particularly towards the halo of the
Galaxy, since the Faraday rotation expected is often much lower than
those located towards the plane.  For the 650 pulsars with rotation
measure data, 284 of these are located over 300 pc above or below the
galactic plane and have a median rotation measure of $\sim$39 rad
m$^{-2}$.  Ionospheric Faraday depth can thus contribute significantly
to the total observed Faraday depth, especially at times near solar
maximum or when observing from lower latitude sites.

The TEC maps from ROB will help to perform differential Faraday
rotation studies between LOFAR core stations and the international
stations. Fig.~\ref{euTECmap} shows a VTEC map from ROB over
Europe. It is clear that the international stations are subject to
different ionospheric conditions than the core, which results in
different amounts of Faraday rotation for the signals arriving at each
station. This effect needs to be calibrated for before combining all
the LOFAR stations to carry out polarization studies.  On the other
hand, to better understand the ionospheric variations above just the
Dutch LOFAR stations (max. baseline $\lesssim$ 100 km), better
geographically resolved TEC maps are needed.  Alternatively, raw,
dual-frequency GPS data should be directly analyzed.  Raw GPS data is
also desirable to use when short timescale ionospheric changes (on the
order of seconds to a few minutes) need to be studied.  For arrays
located in Europe this is possible due to the considerable number of
GPS stations from the permanent European network. However, this is not
yet the case for arrays like the GMRT or future SKA. For instance, in
India there is only one active station
\footnote{See {\tt http://igscb.jpl.nasa.gov/network/hourly.html}}
(near Bangalore) which is part of the IGS network. This is also the
case for South Africa and Western Australia, where there are a few
more stations but they are very far apart. Regional dedicated networks, 
like in Europe, are needed for these arrays in order to gain a more 
detailed picture of the ionospheric sky over these sites. Alternatively,
 high accurate ionospheric corrections for the SKA 
could be possible through its own imaging data; a route being explored, as 
previously mentioned, for LOFAR.

A single LBA station operating at 20 MHz has a beam width of $\sim$
13$^{\circ}$, which spans less than 2 cell widths in the ROB TEC
maps. This is because each of the cells (0.5$^{\circ}$) extends $\sim$
60 km at the altitude of the ionospheric thin shell, which corresponds
to a resolution of $\sim$ 7.5$^{\circ}$ in the plane of the sky.  When
using ROB data, this implies that we expect to predict the same
ionospheric Faraday rotation for any source located within a radius of
7.5$^{\circ}$.

We are expanding \texttt{ionFR} to support also the use of TEC data
from the last release of the IRI (IRI-2007 \footnote{See {\tt
    http://omniweb.gsfc.nasa.gov/vitmo/iri\_vitmo.html}}).  
This addition will allow access to TEC data from before 1995.  This
type of data was not included in the analysis herein due to the fact
that no uncertainties are available for the IRI.

\begin{figure}
\resizebox{\hsize}{!}{\includegraphics{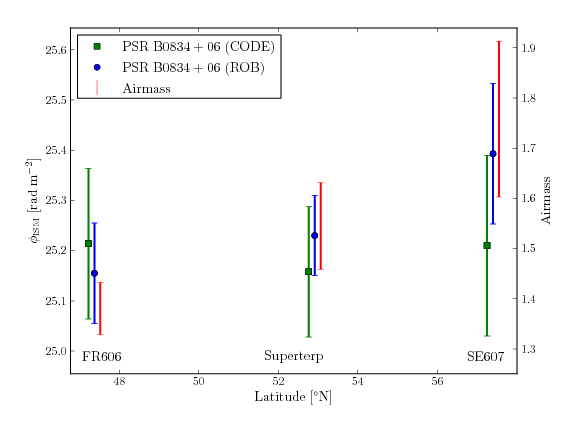}}  

\caption{The measured Faraday depth of the ISM, $\phi_{\mathrm{ISM}}$,
  towards PSR B0834+06, as determined from HBA observations using the
  LOFAR FR606, Superterp and SE607 stations.  These are plotted as a
  function of the geographic latitudes of the stations.  The data
  points are offset slightly in latitude for ease of comparison.  The
  measurements are also compared with the range in airmass calculated
  for the station location towards PSR B0834+06 for the duration of
  the observations (red triangles, {\it right axis}).  The {\tt ionFR}
  model was run using both TEC data from CODE (green squares) and ROB
  (blue points).  }

\label{fig:latRM}
\end{figure}

To further improve the accuracy of the {\tt ionFR}-modeled Faraday depths, 
the Earth will be considered as an ellipsoid and also the various
ionospheric layers (D, E, and F) can be treated.  A three-dimensional
model for the ionosphere should also be possible using ray tracing for
three-dimensional tomography of the ionosphere
\citep[e.g.,]{bath5671}.  An indication that this may be needed for
future high-accuracy Faraday depth determinations with LOFAR is the 
increase in the derived Faraday depth of the ISM towards
B0834+06 as a function of latitude in the international station
observations, see Fig.~\ref{fig:latRM}.  Since the elevation of the
pulsar is lower at higher geographic latitudes in these
quasi-simultaneous observations, the LOS towards B0834+06 passes
through a larger ionospheric depth as seen from Sweden compared with
the Netherlands and France.  Fig.~\ref{fig:latRM} shows how the
 $\phi_{\mathrm{ISM}}$ determined at three latitudes appears
to follow a similar trend to the airmass.  Therefore, a correction
similar to, although not quite as large as, the airmass may be needed
\citep[e.g.,]{1962Natur.195..982W}.

\section{Conclusions}\label{concl}

We have presented {\tt ionFR}, a code that models the ionospheric
Faraday rotation using publicly available TEC maps and the IGRF11.  In
\S3 we show modeled ionospheric Faraday depths for changing levels of
solar activity and different geographic locations.  In \S5 we compare
{\tt ionFR}-modeled Faraday depths with low-frequency data from LOFAR
and the WSRT.  These observational comparisons demonstrate the
robustness and accuracy of the modeled data.

We have shown in \S5.1 that calibrating LOFAR data with {\tt ionFR}
provides very high-precision pulsar RMs (absolute error $\lesssim 0.1$
rad m$^{-2}$).  The applicability of precise RMs is broad.  For
instance, precise RMs can be used to map the structure of the Galactic
magnetic field. Also, it is now possible to monitor pulsar RMs on
multi-year timescales. For example, B0834+06, shows an apparent
increase in its RM after $\sim$ 25 years, assuming previous
ionospheric calibration was correct.

Our code represents an alternative, and moreover, a cheaper solution
when no GPS receivers are co-located with the radio telescope carrying
out the observation. Additionally, GPS receivers may need periodic
maintenance, which requires an investment of time and
money. Therefore, this code is put forward as a simple and costless
method to the community to predict and correct for ionospheric Faraday
rotation.

{\tt ionFR} will be used to correct for the ionospheric Faraday
rotation in projects such as the WSRT Continuum Legacy Survey at 350
MHz: Low-frequency Galaxy Continuum Survey (targets were observed in 2012), 
the LOFAR Magnetism Key Science Project
\citep[MKSP;]{2012arXiv1203.2467A} and the Polarisation Sky Survey
of the Universe's Magnetism \citep[POSSUM;]{2010AAS...21547013G}
project planned with the Australian Kilometre Array Pathfinder 
\citep[ASKAP;]{2007PASA...24..174J} telescope.

Lastly, Phase I of the SKA will provide an unprecedented low-frequency
radio telescope, capable of making, e.g., a detailed map of the
Galactic magnetic field structure through Faraday depth measurements
of pulsars.  However, given that the ionospheric equatorial anomaly
sometimes passes directly over the two sites in South Africa and
Western Australia, it will be crucial to have a robust and accurate
calibration procedure in place to take full advantage of what the SKA
has to offer.

\begin{acknowledgements}

The authors want to express their gratitude to Dr. Dominic Schnitzeler
and Dr. David Champion for their valuable comments and suggestions to
improve the quality of the paper. CSB and CS would like to thank the
Deutsche Forschungsgemeinschaft (DFG) which funded this work within
the research unit FOR 1254 ``Magnetisation of Interstellar and
Intergalactic Media: The Prospects of Low-Frequency Radio
Observations''.  JWTH was funded for this work by a Veni Fellowship of
the Netherlands Foundation for Scientific Research (NWO). OW is
supported by the DFG (Emmy-Noether Grant WU 588/1-1) and by the
European Commission (European Reintegration Grant
PERG02-GA-2007-224897 WIDEMAP). CF and GM acknowledge financial
support by the {\it “Agence Nationale de la Recherche”} through grant
ANR-09-JCJC-0001-01. The Low Frequency Array (LOFAR) was designed and
constructed by ASTRON, the Netherlands Institute for Radio Astronomy,
and has facilities in several countries, that are owned by various
parties (each with their own funding sources), and that are
collectively operated by the International LOFAR Telescope (ILT)
foundation under a joint scientific policy.  The WSRT is operated by
ASTRON/NWO.

\end{acknowledgements}

\bibliographystyle{aa}
\bibliography{ionRM_LOFAR.bib}

\end{document}